\documentclass[aps, pra, reprint, superscriptaddress, nofootinbib,]{revtex4-2}
\usepackage[T1]{fontenc}				
\usepackage[utf8]{inputenc}			
\usepackage[english]{babel}			
\usepackage{lmodern}       
\usepackage[protrusion=false]{microtype}	

\usepackage{xcolor} 						
\usepackage{amsmath} 
\usepackage{amssymb, amsfonts, mathrsfs} 
\usepackage{array, dcolumn} 
\usepackage{graphicx} 
\usepackage{textcomp, gensymb} 
\usepackage{braket}
\usepackage{textgreek}
\usepackage{physics}
\usepackage{bm}
\usepackage{bbold} 
\usepackage{comment}

\definecolor{red}{rgb}{1,0,0}				

\usepackage{soul} 
\definecolor{hlyellow}{rgb}{0.95,0.95,0}
\definecolor{hlgreen}{rgb}{0,0.95,0}
\sethlcolor{hlyellow}
\soulregister \ref 7
\soulregister \cite 7
\soulregister \citet 7
\soulregister \footnote 8
\soulregister {\ } 7
\soulregister \figref 7
\soulregister \tabref 7
\soulregister \secref 7
\soulregister \appref 7
\soulregister \refref 7
\soulregister \eqref 7
\soulregister \supmat 7

\usepackage{hyperref} 
\definecolor{dullmagenta}{rgb}{0.4,0,0.4} 
\definecolor{darkblue}{rgb}{0,0,0.4}
\definecolor{medblue}{rgb}{0,0,0.6}
\definecolor{lightblue}{rgb}{0,0,0.8}
\hypersetup{
	colorlinks=true, 		
	breaklinks=true,		
	linkcolor=lightblue,
	citecolor=lightblue,
	urlcolor=lightblue,
	filecolor=dullmagenta
}
\usepackage[all]{hypcap} 

\newcommand{\figref}[1]{Fig.\ \ref{#1}} 
\newcommand{\tabref}[1]{Tab.\ \ref{#1}} 
\newcommand{\secref}[1]{Sec.\ \ref{#1}} 
\newcommand{\refref}[1]{Ref.\ \cite{#1}} 
\newcommand{\eqnref}[1]{Eq.\ \eqref{#1}} 
\newcommand{\onevsthree}{1 \textit{vs}.\ 3}

\newcommand{\degC}{{}^{\circ} \text{C}} 			
\newcommand{\altket}[1]{| #1 \rangle}      

\newcommand{\Tm}{\ensuremath{\mathrm{T}_{-}(1{,}1)}}
\newcommand{\Tzero}{\ensuremath{\tilde{\mathrm{T}}_{0}(1{,}1)}}
\newcommand{\Szero}{\ensuremath{\tilde{\mathrm{S}}(1{,}1)}}
\newcommand{\oneone}{\ensuremath{(1{,}1)}}
\newcommand{\zerotwo}{\ensuremath{(2{,}0)}}
\newcommand{\zeroone}{\ensuremath{(1{,}0)}}

\newcommand{\Szerotwo}{\ensuremath{\mathrm{S}(2{,}0)}}

\newcommand{\Tp}{\ensuremath{\mathrm{T}_{+}(1{,}1)}}

\newcommand{\supmat}{Supplementary information} 
\newcommand{\methodsref}[1]{Methods \secref{#1}} 

\begin{document}

\newcommand{\zrl}{IBM Research Europe -- Zurich, Säumerstrasse 4, 8803 Rüschlikon, Switzerland}
\newcommand{\yorktown}{IBM Quantum, T.J.\ Watson Research Center, Yorktown Heights, NY, USA}

\title{Identification and mitigation of errors in hole spin qubit readout}%

\author{Eoin Gerard \surname{Kelly}}
	\affiliation{\zrl}
\author{Leonardo \surname{Massai}}
	\affiliation{\zrl}
\author{Bence \surname{Het\'enyi}}
	\affiliation{\zrl}
\author{Marta \surname{Pita-Vidal}}
	\affiliation{\zrl}
\author{Alexei \surname{Orekhov}}
	\affiliation{\zrl}
\author{Cornelius \surname{Carlsson}}
	\affiliation{\zrl}
\author{Inga \surname{Seidler}}
	\affiliation{\zrl}
\author{Konstantinos \surname{Tsoukalas}}
	\affiliation{\zrl}
\author{Lisa \surname{Sommer}}
	\affiliation{\zrl}
\author{Michele \surname{Aldeghi}}
	\affiliation{\zrl}
\author{Stephen~W.~\surname{Bedell}}
	\affiliation{\yorktown}
\author{Stephan \surname{Paredes}}
	\affiliation{\zrl}
\author{Felix~J.\ \surname{Schupp}}
	\affiliation{\zrl}
\author{Matthias \surname{Mergenthaler}}
	\affiliation{\zrl}
\author{Andreas \surname{Fuhrer}}
	\affiliation{\zrl}
\author{Gian \surname{Salis}}
	\affiliation{\zrl}
\author{Patrick \surname{Harvey-Collard}}
	\email{Corresponding author: phc@zurich.ibm.com}
	\affiliation{\zrl}

\date{May 19, 2026}

\begin{abstract}
High-fidelity readout of spin qubits in semiconductor quantum dots can be achieved by combining a radiofrequency charge sensor with spin-to-charge conversion and Pauli spin blockade. However, reaching high readout fidelities with hole spin qubits remains challenging due to site-dependent spin anisotropies and short relaxation times during readout. Here, we analyze the multiple error processes that arise during readout using a double-latched scheme in a germanium double quantum dot hole spin qubit system. We first investigate the spin-to-charge conversion process as a function of magnetic field orientation, and configure the system to adiabatically map the $\lvert \downarrow\downarrow \rangle$ state to the only non-blockaded state. We reveal a strong magnetic field strength dependence of the spin relaxation rates, and minimize it by operating at low fields. We further characterize and mitigate the error processes that arise during the double-latching process. By combining a radiofrequency charge sensor, a double-latching process, and optimized magnetic field parameters, we achieve a single-shot single-qubit state-preparation-and-measurement fidelity of 97.0\% while maintaining control of both spins under the readout conditions.
\end{abstract}

\maketitle

\section{Introduction}

Hole spins in germanium quantum dots (QDs) are a promising spin qubit platform for a variety of reasons: high-quality heterostructures leading to low charge noise and disorder \cite{lodari_low_2021, massai_impact_2024}, all-electrical manipulation due to strong spin-orbit interaction \cite{hendrickx_single-hole_2020}, operational sweet-spots for qubit coherence \cite{hendrickx_sweet-spot_2024} and high-fidelity qubit operations \cite{wang_operating_2024, Lawrie2023}. Spin readout in these systems is typically performed using Pauli spin blockade (PSB) to encode spin into charge, followed by charge readout using a charge sensor. Measuring the charge sensor via radio-frequency (RF) reflectometry can provide high sensitivity and a large bandwidth \cite{vigneau_probing_2023}. A latched readout technique \cite{harvey-collard_high-fidelity_2018} has previously been used in germanium \cite{hendrickx_four-qubit_2021} to enhance the readout signal and lifetime of the charge states during the measurement. Furthermore, a double-latching mechanism \cite{kiyama_high-fidelity_2024} for spin qubit readout has recently been shown to increase of the lifetime of metastable latched charge states which are limited by charge co-tunneling.  

However, achieving $>99\%$ fidelity readout of hole spins remains a challenge. Such high-fidelity readout has been demonstrated in electron spin qubit systems based on silicon \cite{Huang_high-fidelity_2024, philips_universal_2022, blumhof_fast_2022, takeda_rapid_2024}, but the highest reported fidelity for holes in germanium lies around 94\% \cite[see Supplementary Materials]{wang_operating_2024}. Increasing the readout fidelity will require a detailed understanding of how strong spin-orbit interaction and g-tensor anisotropies affect spin-to-charge conversion (StCC) and readout errors. 

In this Article, we analyze StCC and the various decay processes that can occur during spin readout in a germanium hole double QD system with an RF charge sensor. After a brief review of the enhanced latching readout,
 we first investigate 
 the Landau-Zener physics of passing through two different singlet-triplet anticrossings induced by site-dependent spin anisotropies. We distinguish three different StCC regimes based on diabaticity with respect to these anticrossings, and access each in turn by varying preparation and readout ramp times. We then investigate the ramp time required to reach a particular regime as a function of the magnetic field orientation, as well as the errors that arise during StCC. 
We furthermore consider spin decay in the PSB window, and find that spin decay times are increased by operating at low field strengths. The errors resulting from spin and charge decay during latched mapping are also investigated. Finally, 
leveraging the understanding acquired about each readout process, we demonstrate an average single-qubit and single-shot state-preparation-and-measurement (SPAM) fidelity of $97.0(5)\%$ using double latching, primarily limited by imperfect state preparation and signal-to-noise ratio (SNR). 

\begin{figure*}[tbp]
\includegraphics[width=\textwidth]{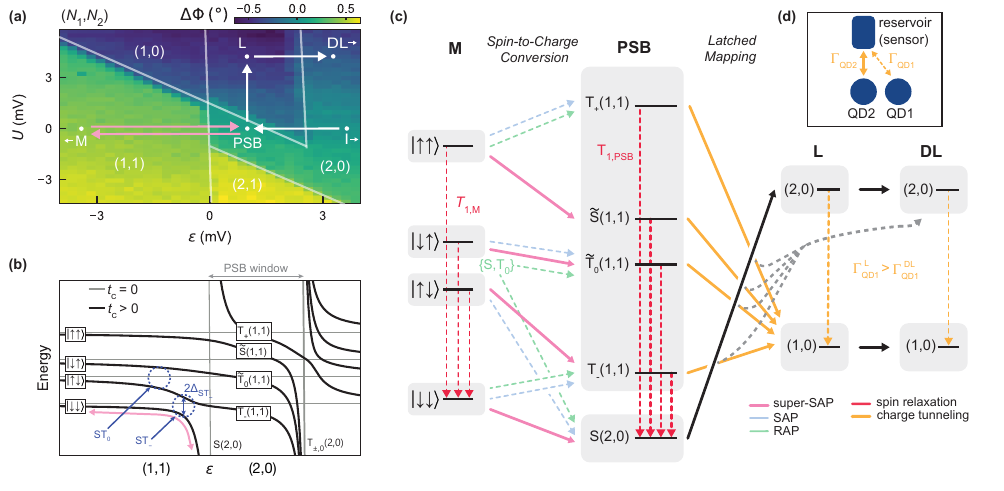}
\caption{ {Double-latched readout processes.} 
\textbf{(a)}~Double quantum dot (QD) charge stability diagram where the virtual detuning $\varepsilon$ and on-site energy $U$ are varied, and the reflectometric change in phase $\Delta\Phi$ is plotted. The points and arrows indicate the pulse sequence. Arrows beside labels indicate points outside the scan area. The pulse sequence also involves the barrier gate vB12 (\figref{fig:fig4}(a)). 
I: initialization point. PSB: point inside the Pauli spin blockade window. M: manipulation point where qubits are operated.  L: latched point, where enhanced latched mapping takes place. DL: double-latched point, used to extend the lifetime of the metastable state after mapping. 
The plot is the result of varying the L point after preparing a blocked state at point M (DL point not used). 
\textbf{(b)}~Simulated two-spin energy spectrum as a function of $\varepsilon$ with $U = 0\,\mathrm{mV}$, with details given in \supmat{} \secref{app:theoretical model}. Here $t_{\mathrm{c}}$ refers to the interdot tunnel coupling. The polarized spin states $\mathrm{T}_{\pm}(1,1)$ hybridize with the singlet $\Szerotwo$  due to a combination of g-factor anisotropy and spin-flip tunneling, leading to an anti-crossing of size $2\Delta_{\mathrm{ST}_-}$.  The location of the ST$_0$ anticrossing (i.e., where the exchange energy equals the Zeeman splitting difference) is also indicated.
\textbf{(c)}~Summary of the conversion processes involved in the readout. Starting from a qubit state at point M, the arrows indicate possible paths that might lead to the desired outcome (full lines) or errors (dashed lines). The diagram can also be used to understand the contribution of each step to the total readout time, including the time to ramp from M to PSB, the time to wait at L, and the charge readout time at DL. Relaxation times $T_1$ contribute to errors.
\textbf{(d)}~Schematic of the connectivity of the double QD system, where QD-reservoir transition rates denoted by $\Gamma$ must be largely asymmetric. 
}
\label{fig:fig1}
\end{figure*}

\begin{figure*}[tbp]
\includegraphics[width=\textwidth]{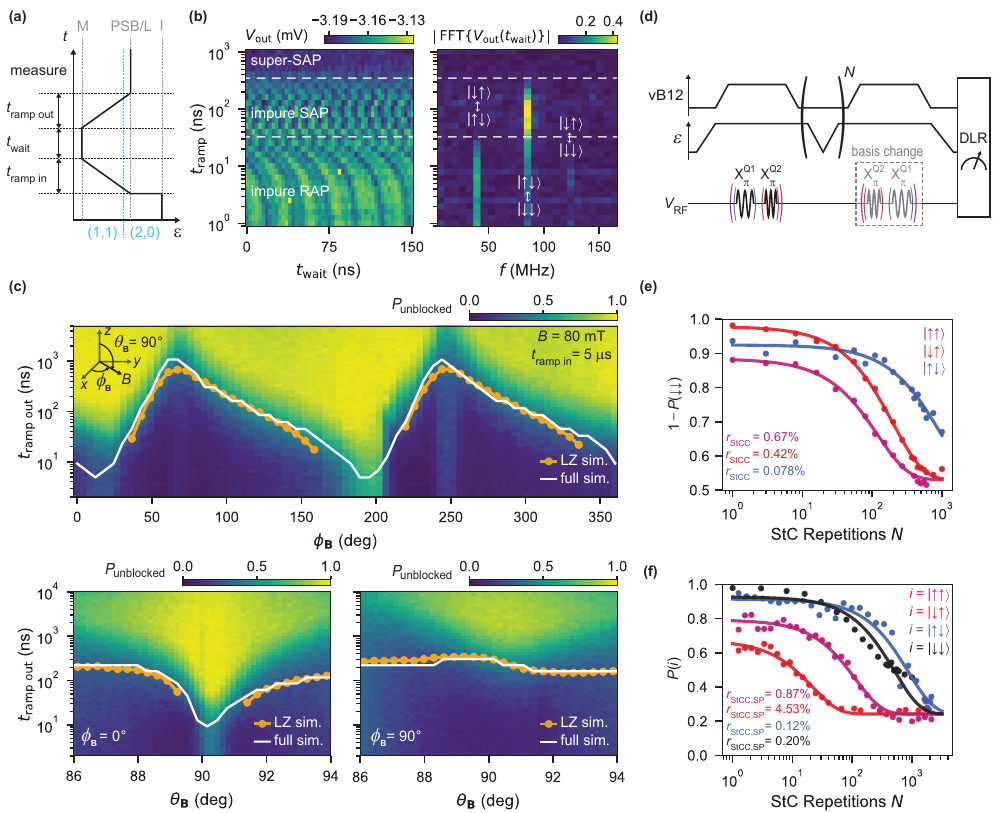}
\caption{ {Spin-to-charge conversion.} 
\textbf{(a)}~Pulsing sequence associated with the investigation of the different spin-to-charge conversion (StCC) regimes. 
\textbf{(b)}~Variation of $t_{\textrm{ramp}}$ = $t_{\textrm{ramp in}}$ = $t_{\textrm{ramp out}}$ versus $t_{\textrm{wait}}$, together with the fast Fourier transform along $t_{\textrm{wait}}$ of the reflectometry voltage $V_\text{out}$. Different transition frequencies are revealed in different StCC regimes, namely super-slow adiabatic passage (super-SAP), SAP and rapid adiabatic passage (RAP). 
\textbf{(c)}~Variation of $t_{\textrm{ramp out}}$ for a fixed $t_{\textrm{ramp in}}=5$\,\textmu s as a function of magnetic field orientation in three different planes with $B = 80$\,mT. $P_\text{unblocked}$ is the fraction of unblocked measurements. The orange and white lines represent the adiabatic-diabatic crossover simulated using the Landau-Zener formula or a time-evolution of the system Hamiltonian, as detailed in the main text. 
\textbf{(d)}~The pulsing sequence used to determine the StCC errors in the super-SAP regime of (e) and (f). A specific $(1{,}1)$ spin eigenstate in \{$\ket{\uparrow \downarrow}, \ket{\downarrow \uparrow}, \ket{\uparrow \uparrow}$\} is first prepared (blue, red and magenta pulses, respectively), followed by a symmetric StCC process repeated $N$ times, before a final StCC step and double-latched readout (DLR). $P(i)$ represents the fraction of shots where the two spins were assigned the state $\ket{i}$. 
\textbf{(e)}~Without the final basis change $\pi$-pulses $\mathrm{X}_\pi$, the protocol extracts the StCC error per double ramp $r_{\mathrm{StCC}}$ (fitted lines) for different input states (legend). This contributes errors in the normal sense of PSB readout.
\textbf{(f)}~With the final basis change pulses, the protocol extracts the probability per double ramp $r_{\mathrm{StCC,SP}}$ that the input state leaks to any other state. This process is not necessarily a readout error if the mixing occurs within the blocked subspace. Both (e) and (f) are measured with $\left(\phi_{\textbf{B}},\theta_{\textbf{B}}\right)=\left(194\degree, 90\degree\right)$. }
\label{fig:fig2}
\end{figure*}

\section{Double-Latched Readout Scheme}
\label{sec:latchedexplanation}

The device used in this experiment is a linear six QD array with two charge sensors \cite{tsoukalas_dressed_2025} created on a Ge/SiGe heterostructure containing a strained Ge quantum well. The device details are reported in \methodsref{app:device_and_exp_setup}. Two single-hole QDs at one end of the array, QD1 and QD2, are formed underneath plunger gates P1 and P2 and are capacitively coupled to a nearby charge sensor. The interdot tunnel coupling between the two QDs is tuned by a barrier gate B12. An LC matching circuit utilising a niobium nitride (NbN) superconducting nanowire inductor on a separate chip \cite{kelly_capacitive_2023, eggli_all-electrical_2024} is used to detect changes in impedance of the charge sensor via RF reflectometry \cite{vigneau_probing_2023}. 

A grounded screening layer in the first gate layer of the device is used to shield the sensor RF excitation from undesired dissipation in the heterostructure. On the one hand, this ensures signal collection from the charge sensor, but on the other hand forces a trade-off with charge sensitivity, which is negatively impacted by the $C_{\mathrm{p}} \approx 6$~pF parasitic capacitance introduced (see \supmat{} \secref{app:rf_matching}). The measurement bandwidth could be significantly improved in the future by using a heterostructure with less dissipation, allowing to forego the screening layer.

Due to the short relaxation time of blocked spin states, a double-latched readout scheme \cite{kiyama_high-fidelity_2024} is used in this experiment to enhance the readout signal's lifetime \cite{harvey-collard_high-fidelity_2018}. The scheme is illustrated in \figref{fig:fig1}(a), where the charge stability diagram around the $\oneone$ to $\zerotwo$ charge sectors is plotted along with the pulse sequence. The measured quantity is the change in phase $\Delta\Phi$ of the reflectometry signal. Here $(N_1, N_2)$ represents the hole charge state of QD1 and QD2, respectively. The device is tuned to load holes from the charge sensor as indicated in \figref{fig:fig1}(d), which  naturally provides the asymmetry in loading rates required for latched readout. The applied gate voltages are parameterised by the on-site energy $U$ and detuning $\varepsilon$ virtual control parameters, which are linear combinations of virtual gate voltages vP1 and vP2 (see \methodsref{app:virtual_gates_detuning}).

The complete readout process involves a sequence of steps with many possibilities for errors to occur, and is comprehensively described in \figref{fig:fig1}(c).  StCC of computational states is first performed by ramping $\varepsilon$ from the manipulation point M to point PSB within the PSB window. The barrier is also changed to increase the tunnel coupling prior to this ramp (\figref{fig:fig4}(a)). This maps the states to either the unblocked $\Szerotwo$ state or blocked $\{\Tm$, $\Tzero$, $\Szero$, $\Tp\}$ states at point PSB. Here the tilde represents the hybridization of ${\mathrm{T}}_{0}(1,1)$ and ${\mathrm{S}}(1,1)$ due to the interplay between the g-factor difference and exchange inside the PSB window. Furthermore, throughout the main text, we take the arrow symbol in the kets like $\ket{\downarrow \downarrow}, \ket{\uparrow \uparrow}$ to represent the ground state and excited state indices, respectively, instead of representing the direction of the spin in real space. This is because the spins could point in very different directions in the two QDs.

Depending on the adiabaticity of the ramp with respect to both the ST$_{-}$ and ST$_{0}$ anticrossings, three different StCC schemes are possible. Here it is assumed that the ramp is always adiabatic with respect to charge. Super-slow adiabatic passage (super-SAP) is adiabatic with respect to both anticrossings, resulting in a mapping depicted by the pink arrows in \figref{fig:fig1}(c). Slow adiabatic passage (SAP) is adiabatic with respect to the $\mathrm{ST}_{0}$ anticrossing, but diabatic with respect to the $\mathrm{ST}_{-}$ anticrossing, resulting in a mapping depicted by the blue arrows \cite{taylor_relaxation_2007}. Rapid adiabatic passage (RAP) is diabatic with respect to both anticrossings (green arrows). A table summarising these different regimes is given in \methodsref{app:stc_mapping_schemes}. 

Operating with super-SAP allows for a state-selective StCC of $\ket{\downarrow \downarrow}$ to $\Szerotwo$, with all other computational states being converted to $\oneone$ states within the PSB window. We later refer to this generalisation of the singlet-triplet readout as the \onevsthree{} readout (\methodsref{app:stc_mapping_schemes}). As we will show, this is the optimal StCC regime to use in systems with a large $\mathrm{ST}_{-}$ anticrossing. Adiabaticity takes care of preserving the weight of the eigenstates during StCC, alleviating the need to monitor the direction of the quantization axis of each QD.

For initialization, the symmetry of the StCC process is leveraged. A singlet $\Szerotwo$ state is first prepared by waiting approximately 10\,\textmu$\mathrm{s}$ at point I, deep in the $\zerotwo$ region. The $\ket{\downarrow\downarrow}$ state is then prepared by adiabatically ramping $\varepsilon$ through the ST$_{-}$ anticrossing as depicted by the pink arrow in \figref{fig:fig1}(b). This ramp is performed with a large exchange energy $J$ (and therefore tunnel coupling) to facilitate StCC. However, this exchange couples the two spins too much for independent single spin operation. Therefore, $J$ is then adiabatically reduced to $J/h < 1$\,MHz using the virtual vB12 barrier gate before qubit operation (see \figref{fig:fig4}(a)). Any other computational spin state $\{\ket{\uparrow\downarrow}, \ket{\downarrow\uparrow}, \ket{\uparrow\uparrow}\}$ can then be prepared by the application of RF control pulses on gates P1 and P2. 

After qubit operation and subsequent StCC, the blocked $\oneone$ states can decay rapidly to $\Szerotwo$ at the PSB point due to spin-orbit interaction \cite{danon_pauli_2009}. This limits the readout fidelity if the required measurement time is  significant with respect to the spin decay time. This decay can be avoided by quickly mapping the spin state to a  much more stable charge state via a latching process \cite{harvey-collard_high-fidelity_2018}. To enable this, we choose asymmetric tunnel rates $\Gamma_{\mathrm{QD}i}$ of each dot $i$ to the reservoir, such that $\Gamma_{\mathrm{QD2}}$ is much faster than the PSB decay rate, while $\Gamma_{\mathrm{QD1}}$ is much longer than the measurement time.
By pulsing from point PSB to the latching point L, the blocked $\oneone$ spin state is thereby quickly mapped to $\zeroone$ with rate $\Gamma_{\mathrm{QD2}}$, while the $\Szerotwo$ state remains as $\zerotwo$. This scheme avoids spin decay of the $\oneone$ spin states. Using a latched readout scheme also has the benefit of enhancing the contrast between the two states because they now differ by a total charge, compared to a change in charge distribution in the normal PSB case \cite{studenikin_enhanced_2012}.

The locked $\zerotwo$ metastable charge state can now decay to $\zeroone$ by co-tunneling. 
In order to further increase its lifetime, an extra step to a double-latched point DL positioned at a large detuning is used. This reduces co-tunneling through the other dot. As will be shown, reading out at DL greatly enhances the lifetime of the metastable $\zerotwo$ charge state, and is highly effective in preventing further errors before readout. However, since DL is outside the PSB window, incomplete conversion at L by not waiting long enough there before DL can lead to errors (grey dashed lines, \figref{fig:fig1}(c)).

\section{Spin-to-Charge Conversion Errors}
\label{sec:stccerrors}

\begin{figure*}[tbp]
\includegraphics[width=\textwidth]{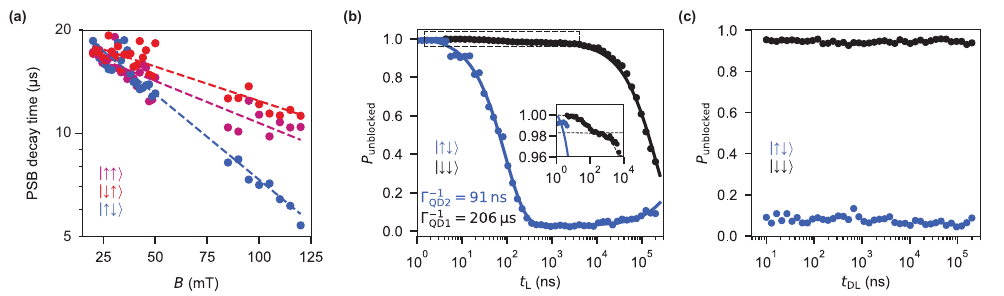}
\caption{ {Readout decay processes.}   
\textbf{(a)}~Characteristic spin decay times $\Gamma_{\mathrm{PSB}}^{-1}$ at the PSB point of the three excited spin eigenstates as a function of magnetic field amplitude $B$. Dashed lines are exponential decay fits to the data. See \methodsref{app:psb_decay_vs_angle} for the angular dependence.
\textbf{(b)}~Probability of measuring the unblocked $\zerotwo$ charge state $P_\text{unblocked}$ as a function of time $t_{\mathrm{L}}$ at the L point for two different input states (blue and black). The inset shows the two-step exponential decay that occurs for $\ket{\downarrow\downarrow}$ within the dashed box. Lines are two-step exponential decay fits to the data. 
\textbf{(c)}~$P_\text{unblocked}$ as a function of wait time $t_{\mathrm{DL}}$ at point DL before measurement for two different spin states ($t_{\mathrm{L}} = 2$\,\textmu s).}
\label{fig:fig3}
\end{figure*}

We now experimentally study the best StCC regime for state preparation and readout.  We first initialize a $\Szerotwo$ state and then pulse $\varepsilon$ from PSB to M. We vary the ramp times to ($t_{\textrm{ramp in}}$) and from ($t_{\textrm{ramp out}}$) the M point, together with the wait time ($t_{\mathrm{wait}}$) at M. The resulting reflectometry signal $V_{\mathrm{out}}$ is shown in \figref{fig:fig2}(b). We set the magnetic field $B = 40\,\mathrm{mT}$ parallel to the QD array, with equal $t_{\textrm{ramp}} = t_{\textrm{ramp in}} = t_{\textrm{ramp out}}$ and plot the results in \figref{fig:fig2}(b). 

For the shortest $t_{\textrm{ramp}}$, we first observe an impure RAP regime (see \methodsref{app:stc_mapping_schemes}). 
Coherent oscillations are observed in the readout signal as a function of $t_{\mathrm{wait}}$ at M \cite{jirovec_dynamics_2022}, as the quick ramp from $\Szerotwo$ into the $\oneone$ charge region prepares a superposition of eigenstates at M. Applying a fast Fourier transform to the data extracts three different oscillation frequencies, as displayed in \figref{fig:fig2}(b) (right panel). These frequencies correspond to all possible transitions between the $\{\ket{\downarrow \downarrow},\ket{\uparrow \downarrow}, \ket{\downarrow \uparrow}$\} states. 

As $t_{\mathrm{ramp}}$ is increased, coherent oscillations between $\ket{\downarrow \downarrow} \leftrightarrow \ket{\uparrow \downarrow}$ become dominant.
This happens when the ramp rate is completely adiabatic with respect to the ST$_{0}$ anticrossing.  Superpositions with respect to the ST$_{-}$ anticrossing are now dominant (impure SAP, \figref{fig:fig2}(b)). 
As $t_{\mathrm{ramp}}$ is increased even further, no coherent oscillations are observed, indicating that the super-SAP regime has been achieved. Here, $\ket{\downarrow \downarrow}$ is mapped to $\Szerotwo$ by adiabatically traversing the $\textrm{ST}_{-}$ anticrossing. The different StCC maps are summarized in \methodsref{app:stc_mapping_schemes} and \figref{fig:fig1}(c). 

At this field orientation, we find no region without coherent oscillations between the RAP and SAP regimes, as evident from \figref{fig:fig2}(b). In other words, there is no ramp time which is completely adiabatic with respect to the $\textrm{ST}_{0}$ anticrossing while being completely diabatic with respect to the $\textrm{ST}_{-}$ anticrossing. This is due to the large ST$_{-}$ anticrossing. 
Previously, a variable ramp rate (double ramp) \cite{petta_coherent_2005} has been used to traverse the two anticrossings at different rates.
In our case, since the regions in detuning where the two anticrossings occur tend to overlap (\figref{fig:fig1}(b)), this scheme is not practical.  
We therefore conclude that in our case, the super-SAP scheme is more suitable. However, when the $\textrm{ST}_{-}$ gap is much smaller than the $\textrm{ST}_{0}$ gap, often the SAP regime is preferable \cite{harvey-collard_all-electrical_2017}. Given this possibility to switch between SAP and super-SAP, the time required by the StCC step should not limit the total readout time until the charge detectors can achieve low integration times of order $100\,\text{ns}$ to $1\,\text{\textmu{}s}$.

Since optimal points for qubit coherence exist at certain magnetic field orientations due to the highly anisotropic nature of hole g-tensors in Ge \cite{hendrickx_sweet-spot_2024}, we investigate the ramp time $t_{\mathrm{ssap}}$ required to reach the super-SAP regime as  a function of field orientation. A method for finding $t_{\mathrm{ssap}}$ is to prepare $\ket{\downarrow \downarrow}$ by using a fixed, large $t_{\mathrm{ramp\,\,in}}$ and vary $t_{\mathrm{ramp\,\,out}}$, while measuring the resulting unblocked state probability $P_{\mathrm{unblocked}}(t_{\mathrm{ramp\,\,out}})$. Such measurements are reported in \figref{fig:fig2}(c) as a function of the azimuthal and polar angles of the magnetic field $\left( \phi_{\textbf{B}},  \theta_{\textbf{B}}\right)$. Here $t_{\mathrm{ssap}}$ is defined such that $P_{\mathrm{unblocked}}(t_{\mathrm{ssap}}) = 1-1/e$. This time varies by orders of magnitude depending on the field orientation due to the modulation of the size of the $\mathrm{ST}_{-}$ anticrossing with field direction. $P_{\mathrm{unblocked}}$ can be modelled by the Landau-Zener (LZ) probability of adiabatically transitioning through the $\mathrm{ST}_{-}$  anticrossing in cases where this anticrossing is well defined \cite{nichol2015a}:
\begin{align}
    P_{\mathrm{LZ}}(t_{\textrm{ramp}}) = 1 - \exp(-\frac{t_{\textrm{ramp}}(\pi \Delta_{\mathrm{ST}_-})^2}{h\left|{(\varepsilon_{\mathrm{PSB}} - \varepsilon_{\mathrm{M}}})\dv{E_{\mathrm{ST_{-}}}}{\varepsilon}\right|}) .
    \label{eq:landau-zener}
\end{align}

We determine $t_{\mathrm{ssap}}$ for a specific magnetic field $\textbf{B}$ by calculating the energy spectrum as a function of $\varepsilon$, and then finding both the size of the anticrossing $\Delta_{\mathrm{ST}_-}$ and slope $\dv{E_{\mathrm{ST_{-}}}}{\varepsilon}$ of the ST$_{-}$ energy difference versus $\varepsilon$ around the anticrossing. This requires the determination of the hole g-tensors of the two QDs, which are presented in \methodsref{app:g-tensors}, and a spin-flip tunneling contribution (explained below). The calculated $t_{\mathrm{ssap}}$ from Eq.\,\,\ref{eq:landau-zener} is in agreement with the data for most field orientations (orange dots, \figref{fig:fig2}(c)). The LZ approximation breaks down however when the anticrossing is comparable to the qubit energy splitting, leading to no well defined minimum in the ST$_{-}$ energy difference.

To model the data for all field angles, we use a time-dependent five-state Hamiltonian to propagate the $\ket{\downarrow\downarrow}$ state as a function of time and determine the probability of $\ket{\downarrow\downarrow}$ evolving into a $\Szerotwo$ state for various ramp times. The result of such a simulation is presented as the white lines in \figref{fig:fig2}(c). In addition to knowledge of the qubit g-tensors, a spin-flip tunneling term \cite{geyer_anisotropic_2024, saez-mollejo_exchange_2024} is required to fit the data (see \supmat{} \secref{app:ramp simulation}). The fit allows the extraction of this spin-flip tunneling term. We find the smallest $t_{\mathrm{ssap}}$ at $\left(\phi_{\textbf{B}},\theta_{\textbf{B}}\right) = \left(194\degree, 90\degree \right)$, and operate at this orientation to minimise spin decay errors during the passage through the ST$_{-}$ anticrossing. 
We find empirically that the range of in-plane magnetic fields where super-SAP can be achieved is quite large in the Ge devices we have worked with. If one sets an arbitrary threshold of a 1\,\textmu{}s adiabatic ramp, nearly all angles except two spots spanning $\sim10$ to 20 degrees are suitable (\figref{fig:fig2}(c)).

To measure the errors that arise during the StCC ramps, we repeat a symmetric StCC step an increasing number of times $N$ before readout  (\figref{fig:fig2}(d)). All four two-qubit basis states are prepared in turn by super-SAP preparation of $\ket{\downarrow \downarrow}$ with $t_{\textrm{ramp in}}=120$\,ns, followed by the application of single spin-flip pulses. 
After this input state preparation, a super-SAP back-and-forth ramp  (M$\rightarrow$PSB$\rightarrow$M) is repeated $N$ times. Following this, the population of the specific input state $\ket{i}$ can be optionally mapped to $\ket{\downarrow \downarrow}$ by further qubit control pulses making the readout selective to that state. We refer to this hereafter as a readout basis change step. Lastly, a final StCC mapping followed by double-latched readout is performed. 

We define the StCC error $r_{\mathrm{StCC}}$ as the probability of not detecting the signal of the prepared state after one back-and-forth ramp. In the absence of a readout basis change step (\figref{fig:fig2}(e)), we obtain an average $r_{\mathrm{StCC}}$ for the three blocked spin states of $0.39\%$ per repetition by fitting the data to an exponential decay $\sim\exp{(-N r_{\mathrm{StCC}})}$ (with scale and offset, valid when $r_\mathrm{StCC} \ll 1$). 
With a readout basis change step, this measurement extracts the state-preserving StCC error $r_{\mathrm{StCC,SP}}$ of state $i$ evolving to \textit{any} other state than itself.   We observe an average $r_{\mathrm{StCC,SP}}$ of 1.43\%, limited primarily by the high $r_{\mathrm{StCC,SP}}$ of $\ket{\downarrow\uparrow}$ (\figref{fig:fig2}(f)). 
Note that errors extracted in this way would not lead to a measurement error when the leakage occurs between the blocked states. It is therefore a more stringent measure that is similar to a quasi-non-demolition measurement of the $\ket{\downarrow\downarrow}$ \textit{vs}.\ $\{ \ket{\downarrow\uparrow}, \ket{\uparrow\downarrow}, \ket{\uparrow\uparrow} \}$ two-spin subspaces, in the sense that it potentially disregards phase coherence but preserves eigenstate probabilities \cite{nurizzo2023}.

\section{Readout Decay Processes}
\label{sec:psbdecayerrors}

\begin{figure*}[tbp]
\includegraphics[width=\textwidth]{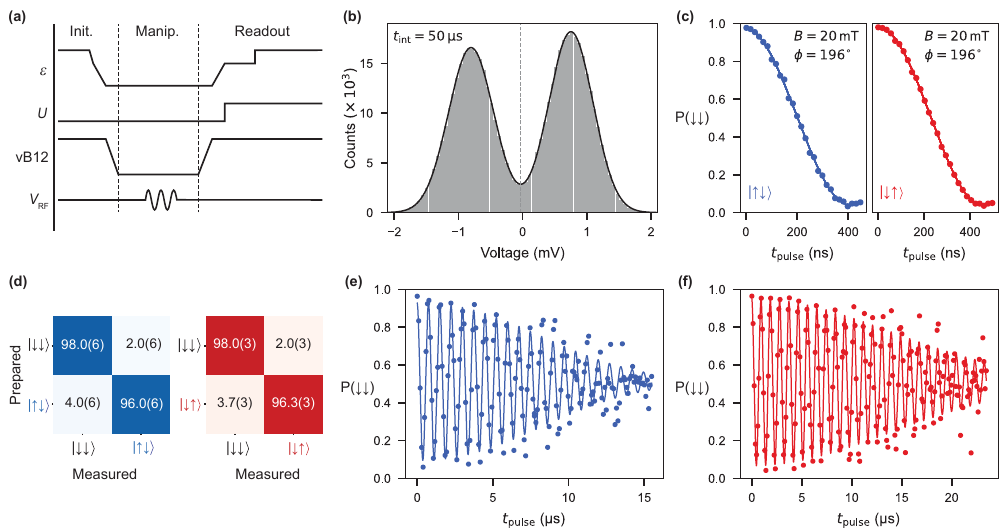}
\caption{ {Single-qubit state preparation and measurement (SPAM) fidelities.} 
\textbf{(a)}~Complete pulse sequence used in this experiment, separated in abstract steps ``initialization'', ``manipulation'', ``readout''.
\textbf{(b)}~Histogram of all single-shot measurements for the first half-period Rabi oscillation of qubit 1 for an integration time $t_\text{int}$ of 50\,\textmu s. The threshold used for each pulse time to obtain single-shot readout fidelities is set to the midway point between the means of two fitted Gaussian distributions. 
\textbf{(c)}~Detailed Rabi oscillations (pulse time $t_\text{pulse}$) between $\ket{\downarrow \downarrow}$ and $\ket{\uparrow \downarrow}$ (blue) and  between $\ket{\downarrow \downarrow}$ and $\ket{\downarrow \uparrow}$ (red) with a fit to extract the SPAM fidelities for each qubit. $P(\ket{\downarrow\downarrow})$ is the fraction of shots assigned to the $\ket{\downarrow\downarrow}$ state.
\textbf{(d)}~The confusion matrices associated with the fits in (c) in percentage. The average SPAM fidelity is $97.0(5)\%$. 
\textbf{(e-f)}~Rabi oscillations for Q1 and Q2 and their decay envelope fits, giving Rabi decay times of $T_2^{\mathrm{R}} = 10.8(7)$\,\textmu s for (e) and $T_2^{\mathrm{R}} = 19.6(13)$\,\textmu s for (f). The digits in parenthesis represent the $1\sigma$ confidence interval.}
\label{fig:fig4}
\end{figure*}

As outlined in \figref{fig:fig1}(c), there are several decay processes that affect readout fidelity. Decay before reaching the PSB point is measured with the method presented in the previous section. We now focus on spin decay at the PSB point and charge mapping errors that occur during latched mapping.

Spin decay times are extracted for the blocked spin states at the PSB point as a function of $B$  with $(\phi_{\textbf{B}}, \theta_{\textbf{B}}) = (196\degree, 90\degree)$, 
and plotted in \figref{fig:fig3}(a). The results reveal an approximately exponential decrease of the decay times for all states with increasing $B$, with all of them starting around 20\,\textmu s at the lowest $B$. This exponential decrease contrasts with the power law behaviour previously predicted and measured for spin qubits \cite{watson_atomically_2017, camenzind_hyperfine-phonon_2018, kornich_phonon-assisted_2018}. Lower lifetimes with increasing $B$ suggests that operating at low $B$ is beneficial not only for qubit coherence, as previously reported for planar Ge \cite{hendrickx_sweet-spot_2024,wang_operating_2024}, but also for readout. Furthermore, the decay times of the polarized and anti-polarized states do not differ much from each other (less than an order of magnitude). This may be due to a long relaxation time of $\Szero$ to $\Szerotwo$ at the PSB point \cite{seedhouse_pauli_2021}. This leads to the singlet-triplet-mode readout observed here which, combined with single-spin rotations, can selectively read any one state in the two-spin Hilbert space from the other three. 
In contrast, the parity-mode readout observed in other works \cite{philips_universal_2022, takeda_rapid_2024} is never observed under our conditions, even for different in-plane field orientations (see \methodsref{app:psb_decay_vs_angle}) and strengths (\figref{fig:fig3}a), and tunnel couplings $\approx 1$\,GHz ($\pm$ factor of $\sim5$).” 

For latched mapping, the time spent at the PSB point can be significantly reduced, but the
decay time at the PSB point $\Gamma_{\mathrm{PSB}}^{-1}$ can still be a prevalent source of error as it is in competition with the fast but finite tunnel rate to the reservoir. We refer to this competition between paths as ``the fork''.  The probability of this decay occurring before latching is given by $\Gamma_{\mathrm{PSB}} / ( \Gamma_{\mathrm{PSB}} + \Gamma_{\mathrm{QD2}} ) $ \cite{harvey-collard_high-fidelity_2018}. $\Gamma_{\mathrm{QD2}}$ is found here by measuring $P_{\mathrm{unblocked}}$ after preparing $\ket{\uparrow \downarrow}$ (\figref{fig:fig3}(b)) as a function of $t_{\mathrm{L}}$, the wait time at L. An exponential decay to $\zeroone$ as a function of $t_{\mathrm{L}}$ occurs with a characteristic time $\Gamma_{\mathrm{QD2}}^{-1}=91$\,ns. In our case, the error arising from the fork is estimated to be around 0.46\%, using the PSB decay time of $\Gamma_{\mathrm{PSB}}^{-1}=20$\,\textmu s observed for the three blocked states at 20\,mT (\figref{fig:fig3}(a)). 
It is worth pointing out that given these short PSB decay times, any sensor would need to be similarly fast ($91 \text{\,ns}$ measurement time) to reach the same low error rates without a latched readout.

The $\ket{\downarrow \downarrow}$ state on the other hand is mapped to the $\zerotwo$ charge state, which is metastable at L (\figref{fig:fig1}(c)). $P_{\mathrm{unblocked}}$ is plotted as a function of $t_{\mathrm{L}}$ for a prepared $\ket{\downarrow \downarrow}$ state in \figref{fig:fig3}(b). A two-step exponential decay fit to the data reveals an initial decay to a probability of 98.3\% due to the imperfect initialization of $\ket{\downarrow \downarrow}$, followed by the main exponential decay due to the metastable $\zerotwo$ charge state relaxing to $\zeroone$ with $\Gamma_{\mathrm{QD1}}^{-1}=206\,$\textmu s.

This metastable decay time can now limit the readout fidelity of $\ket{\downarrow \downarrow}$ when the measurement time $t_{\mathrm{int}}$ is not much faster. The $\zerotwo$ readout error can be estimated as the fraction of shots $1 - e^{-\Gamma_{\mathrm{QD1}} t_{\mathrm{int}}/2}$ that have decayed before $t_{\mathrm{int}}/2$ (single-shot threshold placed midpoint between the two bimodal outcomes). In our case with $t_{\mathrm{int}}=50$\,\textmu s, this error would be 11.4\%. The double-latching technique allows for the reduction of this error. Waiting for an optimal time $t_{\mathrm{L,opt}}$ such that $ \Gamma_{\mathrm{QD2}}^{-1} \ll \,t_{\mathrm{L, opt}} \ll \Gamma_{\mathrm{QD1}}^{-1} $ at point L  before pulsing to DL prevents further decay of the $\zerotwo$ state \cite{kiyama_high-fidelity_2024}. This additional lifetime enhancement comes from the detuning dependence of the metastable charge lifetime: near $\varepsilon=0$ (i.e., in the PSB window), the interdot tunnel coupling hybridizes the double dot and partially lifts the latching condition \cite{harvey-collard_high-fidelity_2018}. In \figref{fig:fig3}(c), we see that using double latching, we do not observe significant decay (for both $\ket{\downarrow \downarrow}$ and $\ket{\uparrow \downarrow}$) for up to 200\,\textmu s. In this case, with $t_{\mathrm{L}} = 2$\,\textmu s, the $\zerotwo$ decay error during integration is reduced to  0.97\%.

\section{Single-Qubit SPAM Fidelity}
\label{sec:spamfidelity}

Now that the relevant readout processes and  mapping errors have been characterized, we combine this learning to maximize the single-shot state-preparation-and-measurement (SPAM) fidelity. The strictest test is to measure the visibility of coherent Rabi oscillations (maximum minus minimum on a probability scale) for each qubit while the other spin is considered as an ancilla. The full pulse sequence is described in \figref{fig:fig4}(a). These measurements are performed at $B=20$\,mT to maximise the decay time at the PSB point. The in-plane field angle is set to $\phi_{\mathbf{B}}=196\degree$ to lower the ramp time required to operate in the super-SAP regime. The transition frequencies of $\ket{\downarrow \downarrow} \leftrightarrow \ket{\uparrow \downarrow}$ (qubit 1) and $\ket{\downarrow \downarrow} \leftrightarrow \ket{\downarrow \uparrow}$ (qubit 2) are 16.61\,MHz and 29.24\, MHz, respectively. The time at the latch point is set to $t_{\mathrm{L}}=770$\,ns. The single-shot $\ket{\downarrow \downarrow}$ probability is assigned by the thresholding of 10,000 shots as described in \methodsref{app:thresholding}. The histogrammed data of all shots obtained up to a half-period Rabi oscillation of qubit 1 is presented in \figref{fig:fig4}(b), with an integration time $t_{\mathrm{int}}$ of 50\,\textmu s giving an SNR of 2.235. 

The SPAM fidelities are reported in \figref{fig:fig4}(d) in the form of a confusion matrix. We find an average SPAM fidelity of 97.0(5)\% over both qubits, defined as the average of the diagonal elements of the confusion matrix. The visibilities for each qubit are 94.0(9)\% and 94.3(4)\%. The digits in parenthesis represent the $1\sigma$ confidence interval throughout this work.
We can estimate the contribution of spin-flip errors using the decay envelope of the Rabi oscillations. Rabi decay times $T_2^{\mathrm{R}}$ of  $10.8(7)$\,\textmu s and $19.6(13)$\,\textmu s are obtained by fitting with the function $e^{-(t/T_2^{\mathrm{R}})^{\alpha}}$ with $\alpha=2$ (\figref{fig:fig4}(e)-(f)). Given a Rabi frequency of 1.1\,MHz and 1.2\,MHz for each qubit, we estimate spin-flip errors of 0.15\% and 0.05\% for $\ket{\uparrow \downarrow}$ and $\ket{\downarrow \uparrow}$, respectively.  
Furthermore, the limited SNR contributes to an error between 1.0\% and 1.3\% for the total average SPAM fidelity as detailed in \methodsref{app:thresholding} which could easily be reduced in future work (mostly by increasing the measurement time). A diagram summarising the probability of each step in the readout scheme is given in \methodsref{app:readout_process_probabilities}.

\section{Conclusions}

We have reported a step-by-step tune-up of hole spin qubit readout with a SPAM fidelity $>97\%$ in Ge using an RF charge sensor and a double-latched readout scheme. The tune-up strategy can also be applied to larger arrays under uniform magnetic field and voltage conditions \cite{seidler2025}, as well as to smaller devices \cite{tsoukalas_dressed_2025}, supporting its robustness.
Super-SAP is chosen as the StCC regime, because it accommodates the relatively large $\Delta_{\mathrm{ST}_-}$ and circumvents the challenges of using double ramps that make SAP difficult in this hole system. Super-SAP maps the $\ket{\downarrow \downarrow}$ state to $\mathrm{S}(2,0)$ while the other blockaded $\ket{\uparrow \downarrow}, \ket{\downarrow \uparrow}, \ket{\uparrow \uparrow}$ states are mapped to (1,1) within the PSB window. 
The ramp time $t_{\mathrm{ssap}}$ required to reach the super-SAP regime varies strongly with magnetic field orientation, and can be modelled accurately from knowledge of the g-tensors and spin-flip tunneling. We operate at a magnetic field orientation that reduces $t_{\mathrm{ssap}}$. However, the compatibility of this orientation with qubit control sweetspots requires further investigation. 

After StCC, the (1,1) lifetimes are limited by spin decay at the PSB point. We find that these are relatively short ($\lesssim 20$\,\textmu{}s), and can be extended by operating at low magnetic field. This relaxation alone would yield an average readout fidelity of 64\% for the PSB readout without latching, assuming the same $t_{\mathrm{int}}=50$\,\textmu s. 
This is the main reason why readout in Ge, and for holes in general, is challenging; it requires either very fast charge detection ($t_{\mathrm{int}}=1$\,\textmu s and 25\,\textmu s lifetime would give 99\% fidelity) or a signal-lifetime-enhancing readout. 
Further extension of the PSB decay time could be explored in future work, such as by tuning of the tunnel couplings, detuning, orbital spacings, out-of-plane field, etc., but might yield only modest improvements. We note that the conditions that lead to a high-fidelity parity-mode readout \cite{hendrickx_four-qubit_2021} could not be reached in our parameter range.
The enhanced latching readout (without double latching) increases this fidelity to a predicted value of 92.5\% under the same measurement conditions by replacing the fast relaxing $\oneone$ spin states with the stable $\zeroone$ state. This readout fidelity is limited to a small extent by the fork, but more so by the metastable nature of the $\zerotwo$ charge state.  
The lifetime of the metastable state is then itself extended by pulsing to a double-latched point at a higher detuning.  With this, we achieve an average single-qubit SPAM fidelity of $97.0(5)\%$, limited by $\ket{\downarrow\downarrow}$ initialization errors (1.7\%) as well as SNR errors ($\lesssim 1.3\%$). Both of these error sources could be reduced in future work, and thus SPAM fidelities above $99\%$ should be possible.  

We note that speed of the enhanced latching readout is fundamentally similar to, or better than, the PSB readout: the steps before the final signal integration are usually shorter than the integration time, and the signal is triggered deterministically at the beginning of the integration window because the reservoir tunneling events are fast, while the signal is typically enhanced by the larger charge response of the sensor.  Our work provides a detailed understanding of readout error processes, which we have used here to optimize the single-shot readout fidelity. Moreover, the approach provides increased  fidelities and does so while maintaining qubit control, which is a crucial step for the implementation of quantum computing protocols on this platform.



\newcounter{extfigure} \setcounter{extfigure}{0} 
\makeatletter
\renewcommand\thefigure{\mbox{Ext.~Data~\arabic{extfigure}}}
\makeatother

\newcounter{exttable} \setcounter{exttable}{0} 
\makeatletter
\renewcommand\thetable{\mbox{Ext.~Data~\arabic{exttable}}}
\makeatother

\newenvironment{supfigure}[1][]{\begin{figure}[#1]\addtocounter{extfigure}{1}}{\end{figure}\ignorespacesafterend}
\newenvironment{supfigure*}[1][]{\begin{figure*}[#1]\addtocounter{extfigure}{1}}{\end{figure*}\ignorespacesafterend}
\newenvironment{suptable}[1][]{\begin{table}[#1]\addtocounter{exttable}{1}}{\end{table}\ignorespacesafterend}
\newenvironment{suptable*}[1][]{\begin{table*}[#1]\addtocounter{exttable}{1}}{\end{table*}\ignorespacesafterend}


\renewcommand{\appendixname}{Methods}
\appendix
\newcommand{\supsection}[1]{\section{#1}} 

\supsection{Device and experimental setup}
\label{app:device_and_exp_setup}

\begin{supfigure}[bp]
\includegraphics[width=\columnwidth]{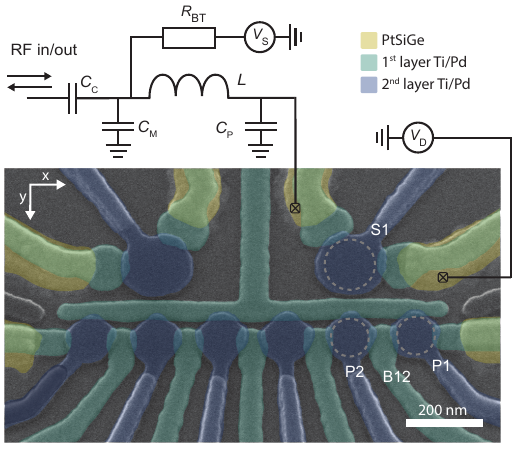}
\caption{ {Device and readout circuit.}
Ge quantum dot (QD) device with RF single-hole transistor, and schematic of the readout circuit (resistors $R$, capacitors $C$, inductors $L$, voltage source $V$). }
\label{fig:fig5}
\end{supfigure}

A false-colored scanning electron microscopy image of the device used in this experiment is presented in \figref{fig:fig5}. The device is fabricated on a Ge/SiGe heterostructure containing a strained Ge quantum well. Ohmic contacts to the quantum well are defined by diffusion of Pt into the upper SiGe barrier at $300\degC$. Two gate layers are then defined on the surface of the heterostructure, each consisting  of SiO$_{2}$ (7 nm) deposited via plasma-enhanced atomic layer deposition followed by Ti/Pd (20 nm) deposited in a lift-off process together with e-beam lithography. Two hole QDs are formed underneath gates P1 and P2, with the exchange coupling being tuned by gate B12. The RF single-hole transistor is formed underneath S1. The device is packaged in 100\,nm SiO$_2$ deposited via plasma-enhanced chemical vapour deposition, with the gates being contacted through vias in the oxide with W interconnects to a bondpad layer. A parallel plate capacitor with a capacitance of $C_{\mathrm{p}} \approx 6\,\mathrm{pF}$ is formed between the bondpad and fanout of each device electrode, together with a screening layer metallization deposited in the first gate layer. The details of the matching circuit are given in \supmat{} \secref{app:rf_matching}.

The measurements are conducted in a Bluefors LD400 dilution refrigerator. The sample chip containing the device together with the NbN nanowire inductor chip are both glued onto a QDevil QBoard circuit board. DC voltages are applied to the device gates via a QDevil QDAC through DC looms inside the fridge that are filtered at the millikelvin stage with a QDevil Qfilter. All plunger gates, barrier gates and the inductor connected to the sensor ohmic are attached to on-PCB bias tees that are connected to attenuated coaxial lines for the application of RF signals. The RF signals are generated using Tektronix AWG5024 arbitrary waveform generators. 

The ohmic of the charge sensor on the right-hand side of the device is connected to a resonant tank circuit in the form of an NbN superconducting nanowire inductor on a separate chiplet with $L=$ 800\,nH, leading to a resonance frequency of 72.7\,MHz. The resonant excitation signal is applied to the resonator only during $t_\text{int}$ using a Quantum Machines {OPX+} ($25 \text{\,mVp}$) via an attenuated line ($-43 \text{\,dB}$) and a directional coupler ($-20 \text{\,dB}$) mounted on the mixing chamber plate, and the reflected signal is amplified by a Cosmic Microwave CMT-BA1 cryogenic amplifier at 4\,K ($+32 \text{\,dB}$) followed by further amplification at room temperature (B\&Z 0.1--3\,GHz, $+37 \text{\,dB}$). 
The readout signal is detected and demodulated by the {OPX+}. The dilution refrigerator is operated at 500\,mK to counteract readout signal shifts attributed to heating of the device that occurs when applying readout pulses to the resonator. This temperature didn't have a noticeable effect on the measured relaxation times.

An American Magnetics three-axis magnet with a maximum field of $(1,1,6)$\,T in the $(x, y, z)$ directions was used to apply the magnetic field and a high-stability current source was used for all coils.

\supsection{Virtual gates, detuning and on-site energy}
\label{app:virtual_gates_detuning}

Virtual gates are used to compensate the cross-capacitance between different physical gates when tuning the electrochemical potentials of the QDs. These are defined in terms of a virtual gate matrix. 

The detuning and on-site energy are defined as linear combinations of virtual gate voltages:
\begin{align}
    \begin{pmatrix}
    \mathrm{vP1} \\
    \mathrm{vP2}
    \end{pmatrix} = 
    \begin{pmatrix}
    -0.5 & 0.5 \\
    0.5 & 0.5
    \end{pmatrix}
    \begin{pmatrix}
    \varepsilon \\
    U
    \end{pmatrix} .
\end{align}

\supsection{Measurement of qubit g-tensors}
\label{app:g-tensors}

The qubit effective g-factors are extracted by determining the resonance frequency as a function of magnetic field orientation at $B=80$\,mT,  as presented in \figref{fig:fig7}. The g-tensor of each qubit is found by a fit to the measured g-factors, giving:
\begin{gather}
\hat{g}_{\mathrm{Q1}} = 
\begin{pmatrix}
    0.046 & -0.075 & -0.038 \\
    -0.075 & 0.384 & -0.007 \\
    -0.038 & -0.007 & 12.4 
\end{pmatrix} ,\\
\hat{g}_{\mathrm{Q2}} = 
\begin{pmatrix}
    0.07 & -0.109 & 0.078 \\
    -0.109 & 0.33 & 0.028 \\
    0.078 & 0.028 & 12.499 
\end{pmatrix} .
\end{gather}\\

\begin{supfigure}[htbp]
\includegraphics[width=\linewidth]{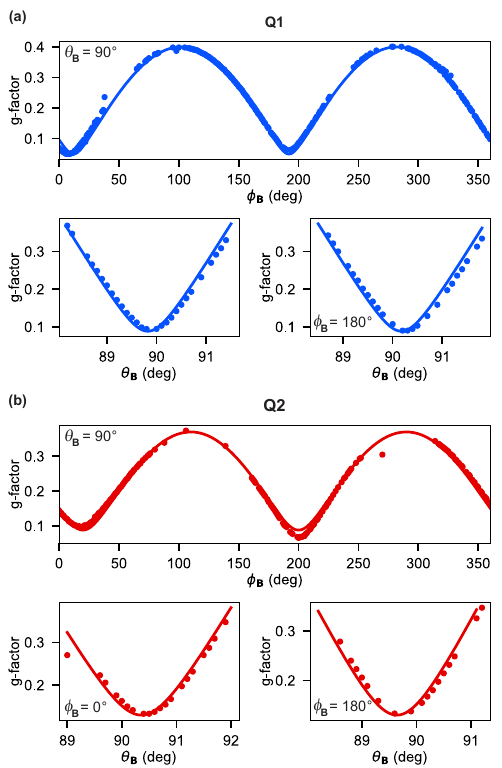}
\caption{Measured g-factors and their fits (lines). 
\textbf{(a)} Q1 and 
\textbf{(b)} Q2 as a function of magnetic field orientation with $B=80$\,mT.}
\label{fig:fig7}
\end{supfigure}

\supsection{PSB decay versus in-plane magnetic field orientation}
\label{app:psb_decay_vs_angle}

\figref{fig:psb_decay_vs_angle} shows the angular dependence of the decay times inside the PSB window. Decay times are extracted by fitting to an exponential (with offsets). The data are not completely compatible because there is also a detuning dependence to these times, and this was not consistently controlled in this dataset. Nevertheless, we do not find special angles where the decay times become very long or very short. Instead, the decay time varies by about a factor two.

\begin{supfigure}[htbp]
\includegraphics[width=\linewidth]{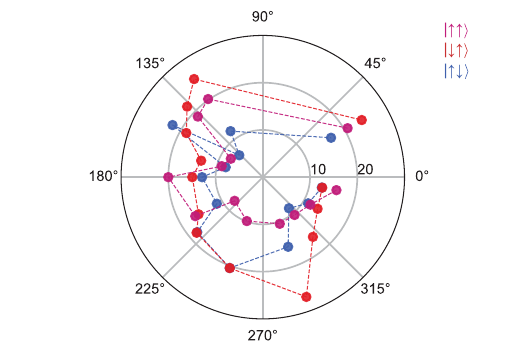}
\caption{Measured decay times (in \textmu s) at the PSB point for $\{ \ket{\uparrow \downarrow}, \ket{\downarrow \uparrow}, \ket{ \uparrow \uparrow}\}$ as a function of in-plane magnetic field angle $\phi_{\mathrm{\textbf{B}}}$ with $\theta_{\mathrm{\textbf{B}}}=90 \degree$ for constant $B=80\,\mathrm{mT}$.
}
\label{fig:psb_decay_vs_angle}
\end{supfigure}

\supsection{Spin-to-charge conversion mapping schemes}
\label{app:stc_mapping_schemes}

\renewcommand{\arraystretch}{1.3}
\setlength{\tabcolsep}{4pt}

Table \ref{table:StC schemes} shows a summary of the different StCC maps. Fully defining the qubit readout in the quantum computing sense requires specifying the qubit states, StCC map, and singlet-triplet or parity-mode PSB regime.
\paragraph{Qubit states}
For the SPAM result, we focus on spin-1/2 qubit encoding, with the other spin used as an ancilla. However, in our study of the decay mechanisms, we also investigate the full two-spin Hilbert space to gain insights and preserve generality.
\paragraph{StCC map}
In this work, we use the super-SAP map, which adiabatically converts the $\ket{\downarrow\downarrow}$ state to a non-blockaded singlet state, while all other three two-spin states remain blockaded.
\paragraph{PSB regime: singlet-triplet or parity-mode}
Fast decay of the blockaded T$_0$-like state can lead to a parity-mode readout (2 \textit{vs}.\ 2), as opposed to singlet-triplet-mode readout (1 \textit{vs}.\ 3).

For all of these steps, the qubit encodings, StCC map and PSB mode can be mixed and matched for achieving a certain result. One must therefore clearly specify the combination used in a given experiment.  

After one measurement, and for \textit{all} the variations, only a single bit of information is acquired, while two bits would be required to fully determine the two-spin state before measurement. One must therefore encode the qubit in a subspace, or repeat quasi-non-demolition measurements in different bases.

\begin{suptable}[htbp]
\centering
\begin{tabular}{llll}
\hline \hline
   &     super-SAP    &    SAP      &     RAP          \\  \hline 
$\ket{\uparrow \uparrow}$ & $\altket{\tilde{\mathrm{S}}(1,1)}$  & $\ket{\mathrm{T}_{+}(1,1)}$   & $\ket{\mathrm{T}_{+}(1,1)}$  \\ \hline
$\ket{\downarrow \uparrow}$ & $\altket{\tilde{\mathrm{T}}_0(1,1)}$  & $\altket{\tilde{\mathrm{T}}_0(1,1)}$  & $\alpha \altket{\tilde{\mathrm{S}}(2,0)}  + \beta \altket{\tilde{\mathrm{T}}_0(1,1)}$ \\ \hline
$\ket{\uparrow \downarrow}$ & $\ket{\mathrm{T}_{-}(1,1)}$ & $\ket{\mathrm{S}(2,0)}$   & $\beta^* \altket{\tilde{\mathrm{S}}(2,0)}  - \alpha^* \altket{\tilde{\mathrm{T}}_0(1,1)}$        \\ \hline
$\ket{\downarrow \downarrow}$ & $\ket{S(2,0)}$   & $\ket{\mathrm{T}_{-}(1,1)}$ & $\ket{\mathrm{T}_{-}(1,1)}$ \\   
\hline \hline
\end{tabular}
\caption{Different StCC schemes for mapping of spin states at point M to spin states at the PSB point. Here $\alpha$ and $\beta$ are complex probability amplitudes with $|\alpha|^2 + |\beta|^2=1$. }
\label{table:StC schemes}
\end{suptable}

\supsection{State probability determination}
\label{app:thresholding}

The $\ket{\downarrow \downarrow}$ single-shot probability is assigned by the thresholding of 10,000 shots. For each measurement, the in-phase and quadrature demodulated signals are projected along one dimension $V_{\mathrm{1D}}$ using principal component analysis, followed by a fit of a bimodal Gaussian distribution to the histogrammed data:
\begin{equation}
C(V_{\mathrm{1D}}) = A_1 e^{\frac{\left(V_{\mathrm{1D}} - \mu_1 - V_{\mathrm{off}}\right)^2}{2\sigma_1^2}} + A_2 e^{\frac{\left(V_{\mathrm{1D}} - \mu_2 - V_{\mathrm{off}}\right)^2 }{2\sigma_2^2}},
\end{equation}
where $C$ is the number of shots for a measured $V_{\mathrm{1D}}$, and $V_{\mathrm{off}}$ is an offset of the bimodal distribution from $V_{\mathrm{1D}}=0\,\mathrm{mV}$. The means $\mu_i$ of each Gaussian $i$ are such that $\mu_1 < \mu_2$. $A_i$ and $\sigma_i$ are the weight and standard deviation of each Gaussian, respectively. We define the signal-to-noise ratio (SNR) as:
\begin{equation}
    \mathrm{SNR} = \frac{|\mu_2 - \mu_1|}{\sigma_1 + \sigma_2}.
\end{equation}
The threshold voltage $V_{\mathrm{th}}$ for single-shot probability assignment is defined as the midway point between the two means, shifted by  $V_{\mathrm{off}}$:
\begin{equation}
    V_{\mathrm{th}} = V_{\mathrm{off}} + \frac{1}{2}(\mu_2 - \mu_1).
\end{equation}
This is valid for $\sigma_1 = \sigma_2$, which is the case here. All shots with $V_{\mathrm{1D}} < V_{\mathrm{th}}$ were assigned as $\ket{\downarrow \downarrow}$.

It is important to distinguish the $\ket{\downarrow \downarrow}$ single-shot probability $P(\ket{\downarrow \downarrow})$ from the averaged assignment probability $A(\ket{\downarrow \downarrow})$, which compares the relative areas of each Gaussian. This is given by ($\sigma_1 = \sigma_2$)
\begin{equation}
    A(\ket{\downarrow \downarrow}) = \frac{A_1}{A_1 + A_2} .
\end{equation}
$A(\ket{\downarrow \downarrow})$ allows for the assignment of the $\ket{\downarrow \downarrow}$ probability once the full bimodal Gaussian distribution is known. This distinguishes readout mapping errors from SNR errors that occur during single-shot thresholding. Calculating the total average assignment fidelity for the measurements of \figref{fig:fig4} gives $98.0(5)\%$. This allows us to assign a $\sim$1\% error in the measured single-shot SPAM fidelity that arises due to SNR limitations. Alternatively, the SNR error $r_{\mathrm{SNR}}$ can be calculated as:
\begin{equation}
r_{\mathrm{SNR}} = \frac{1}{2} \operatorname{erfc}\left(\frac{\mathrm{SNR}}{\sqrt{2}}\right),
\end{equation}
where $\operatorname{erfc}(x)$ is the complementary error function \cite{krantz_quantum_2019}. For the SNR of 2.235 in this experiment, the calculated $r_{\mathrm{SNR}}$ is 1.27\%.

\supsection{Readout process probabilities}
\label{app:readout_process_probabilities}

The probabilities of each step occurring in the double-latched readout scheme are summarised in \figref{fig:fig_readout_errors}. The StCC errors (dashed arrows between M and PSB) are calculated as $1 - \sqrt{1 - r_{\mathrm{StCC}}}$ for the $r_{\mathrm{{StCC}}}$ of each state as presented in the main text. 

\begin{supfigure}[htbp]
\includegraphics[width=\linewidth]{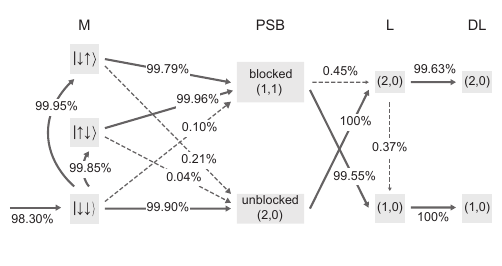}
\caption{Extracted probabilities of individual readout processes. The sequence is to be interpreted as propagating from left to right.}
\label{fig:fig_readout_errors}
\end{supfigure}

\begin{acknowledgments}
We thank the Cleanroom Operations Team of the Binnig
and Rohrer Nanotechnology Center (BRNC) for their help
and support.
\paragraph*{\bf Author contributions}
EGK performed the experiment and analysed the results with help from LM, MPV, and AO. 
BH performed simulations of the super-SAP process. 
EGK, AO, CC and GS developed the RF readout circuit with help from SP. 
SP and LM contributed to development of the experimental setup. 
SWB grew the heterostructure.
FS and MM fabricated the device. 
EGK fabricated the nanowire inductor with help from AO and CC. 
AO, IS, KT, LS and MA contributed to discussions about the experimental data and device improvements.
EGK wrote the manuscript with help from PHC, BH and GS, along with input from all co-authors. 
PHC and GS supervised the project with help from AF.
\paragraph*{\bf Funding}
This project has received funding from the European Union’s Horizon 2020 research and innovation programme under the Marie Skłodowska-Curie grant agreement number 847471, and by the NCCR SPIN under grant numbers {51NF40-180604} and {51NF40-225153} of the Swiss National Science Foundation.
\paragraph*{\bf Competing interests}
The authors declare no competing interests.
\paragraph*{\bf Data availability}
The data that support the findings of this study are available in a Zenodo repository at \url{https://doi.org/10.5281/zenodo.21671988}. Supplementary information is available with this paper.
\end{acknowledgments}

\bibliography{references.bib}

\renewcommand{\appendixname}{Supplementary Section}

\makeatletter
\renewcommand\thesection{\mbox{S\arabic{section}}}
\renewcommand\thesubsection{\mbox{\alph{section}}}
\renewcommand\thesubsubsection{\mbox{\roman{section}}}
\makeatother
\setcounter{section}{0}     

\newcounter{supfigure} \setcounter{supfigure}{0} 
\makeatletter
\renewcommand\thefigure{\mbox{S\arabic{supfigure}}}
\makeatother

\newcounter{suptable} \setcounter{suptable}{0} 
\makeatletter
\renewcommand\thetable{\mbox{S\arabic{suptable}}}
\makeatother

\renewenvironment{supfigure}[1][]{\begin{figure}[#1]\addtocounter{supfigure}{1}}{\end{figure}\ignorespacesafterend}
\renewenvironment{supfigure*}[1][]{\begin{figure*}[#1]\addtocounter{supfigure}{1}}{\end{figure*}\ignorespacesafterend}
\renewenvironment{suptable}[1][]{\begin{table}[#1]\addtocounter{suptable}{1}}{\end{table}\ignorespacesafterend}
\renewenvironment{suptable*}[1][]{\begin{table*}[#1]\addtocounter{suptable}{1}}{\end{table*}\ignorespacesafterend}

\makeatletter
\renewcommand\theequation{\thesection.\arabic{equation}}
\renewcommand\theHequation{S\thesection.\arabic{equation}} 
\makeatother

\onecolumngrid 
\clearpage
{\centering
\large\textbf
{Supplementary information for: \\ Identification and mitigation of errors in hole spin qubit readout} \\ \rule{0pt}{12pt}
}
\twocolumngrid


\supsection{RF matching circuit}
\label{app:rf_matching}

The simplified matching circuit used in this work is presented in \figref{fig:fig5}. A resonant RF signal is coupled into the tank circuit via a coupling capacitor $C_{\mathrm{c}} = 10$\,pF. A spurious matching capacitance of  $C_{\mathrm{m}} = 5$\,pF is necessary to include in the simulation to account for PCB parasitics. The sensor ohmic connected to the inductor is biased through a bias $V_{\mathrm{S}}$ applied through a $R_\text{BT} = 1$\,M$\Omega$ resistor. The parasitic capacitance $C_{\mathrm{p}}$ comes mainly from the parallel plate capacitance between the ohmic bondpad plus fanout and the screening layer below them (the purpose of which was discussed in the main text). The sensor drain ohmic is biased with a voltage $V_{\mathrm{D}}$. All experiments reported in the main text were performed with $V_{\mathrm{S}} = V_{\mathrm{D}} = 0$. 

\begin{supfigure}[bp]
\includegraphics[width=8.4cm]{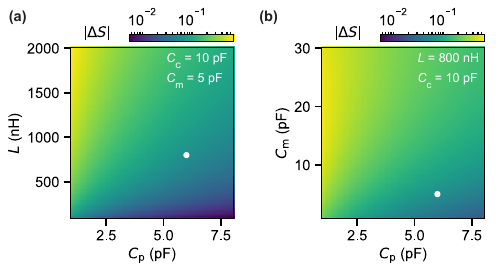}
\caption{Numerical simulations of $S$-parameter difference $|\Delta S|$ for different circuit parameters. 
\textbf{(a)} $C_{\mathrm{p}}$ vs $L$ and 
\textbf{(b)} $C_{\mathrm{p}}$ vs $C_{\mathrm{m}}$. The white points indicate the actual circuit parameters used in this work.}
\label{fig:fig6}
\end{supfigure}

To optimize the charge sensitivity of the matching circuit as a function of different circuit parameters, we defined a measure of sensitivity $|\Delta S|$ as the change in reflected signal when the (sensor) load resistance changes from $100$\,k$\mathrm{\Omega}$ to $1$\,M$\mathrm{\Omega}$. This estimates a realistic change in impedance of a charge sensor in the Coulomb blockade regime. We then numerically simulate for $|\Delta S|$ as a function of $C_{\mathrm{p}}$~vs.~$L$ and $C_{\mathrm{p}}$~vs.~$C_{\mathrm{m}}$ (\figref{fig:fig6}). The white points show the actual parameters used in this work. Reducing the parasitic capacitance $C_{\mathrm{p}}$ to the screening layer, increasing $L$ and also using a higher value of $C_{\mathrm{m}}$ would increase the sensitivity of the RF charge sensor leading to a higher signal-to-noise ratio.  However, increasing $L C_{\mathrm{m}}$ is not feasible due to the reduction in the resonance frequency that would then become attenuated by high-pass filters on the lines, and $C_{\mathrm{p}}$ has already been minimised to the extent allowed by the screening layer. Hence our parameters are an optimum given these constraints.

\phantom{ABCD} 

\supsection{Simulation of the ramp rates}
\label{app:ramp simulation}

The single-particle Hamiltonian of a double QD with spin-orbit interaction reads \cite{geyer_anisotropic_2024}
\begin{align}
    \begin{split}
        H_{1P} &= -\frac \varepsilon 2 (\ket 1\! \bra 1-\ket 2\! \bra 2)\\
        &\quad + \frac {\mu_\text{B}} 2 \mathbf B \cdot (\ket 1\! \bra 1 \hat g_\mathrm{Q1}+\ket 2\! \bra 2 \hat g_\mathrm{Q2}) \boldsymbol \sigma\\
        &\quad + [t_\mathrm{o} \ket 1\! \bra 2 - i t_\mathrm{so} \mathbf n_\mathrm{so} \cdot  \boldsymbol \sigma \ket 1\! \bra 2 +h.c.],
    \end{split}
\label{eq:H1P}
\end{align}
where $\ket i$ is the first orbital state of QD$i$, $\boldsymbol \sigma$ is the spin vector operator with $\sigma_z = \ket \uparrow\! \bra \uparrow - \ket \downarrow\! \bra \downarrow$, $t_\mathrm{o}$ ($t_\mathrm{so}$) is the spin-conserving (spin-flip) tunnel-coupling and $h.c.$ denotes the Hermitian conjugate. Here $\varepsilon$ is the detuning in terms of energy and not voltage as in the main text.

Using the single-particle Hamiltonian from \eqnref{eq:H1P} we construct a particle-exchange-symmetric two-particle Hamiltonian
\begin{multline}
        H_{2P} = H_{1P} \otimes \mathbb{1} + \mathbb{1} \otimes H_{1P} \\
        + U_H (\ket 1\! \bra 1 \otimes \ket 1\! \bra 1  + \ket 2\! \bra 2 \otimes \ket 2\! \bra 2),
    \label{eq:H2P}
\end{multline}
where $U_H$ is the Hubbard charging energy (not the virtual control parameter). Since the
eigenvalues corresponding to particle-exchange-antisymmetric wavefunctions of \eqnref{eq:H2P} obey Pauli's exclusion principle, we project the symmetric Hamiltonian to the following antisymmetric basis states
\begin{subequations}
\begin{gather}
    \ket{S\zerotwo}_0 = \frac 1 {\sqrt 2} (\ket{1\!\uparrow}\otimes \ket{1\!\downarrow} - \ket{1\!\downarrow} \otimes \ket{1\!\uparrow}) , 
    \label{eq:basis5a} \\
    \ket{ss'}_0 = \frac 1 {\sqrt 2} (\ket{1 s} \otimes \ket{2 s'} - \ket{2 s'} \otimes \ket{1 s}) ,
\label{eq:basis5b}
\end{gather} 
\end{subequations}
where we used the $0$ subscript to distinguish the basis states from the eigenstates at the manipulation point $\varepsilon_\mathrm{M}$. 
The projection leads to the low-energy Hamiltonian
\begin{widetext}
\begin{align}
    H_{5\times 5} &= \begin{pmatrix}
        U_H - \varepsilon & t_\mathrm{so}(n_\mathrm{so}^y - i n_\mathrm{so}^x) & t_\mathrm{o} + i t_\mathrm{so}n_\mathrm{so}^z & -t_\mathrm{o} + i t_\mathrm{so}n_\mathrm{so}^z & t_\mathrm{so}(n_\mathrm{so}^y + i n_\mathrm{so}^x)\\
        t_\mathrm{so}(n_\mathrm{so}^y + i n_\mathrm{so}^x) & \Delta_1^z + \Delta_2^z & \Delta_2^x -i \Delta_2^y  & \Delta_1^x -i \Delta_1^y & 0\\
        t_\mathrm{o} -i t_\mathrm{so}n_\mathrm{so}^z & \Delta_2^x +i \Delta_2^y & \Delta_1^z - \Delta_2^z & 0 & \Delta_1^x -i \Delta_1^y\\
        -t_\mathrm{o} - i t_\mathrm{so}n_\mathrm{so}^z & \Delta_1^x +i \Delta_1^y & 0 & -\Delta_1^z + \Delta_2^z & \Delta_2^x - i \Delta_2^y\\
        t_\mathrm{so}(n_\mathrm{so}^y - i n_\mathrm{so}^x) & 0 & \Delta_1^x +i \Delta_1^y & \Delta_2^x +i \Delta_2^y & -\Delta_1^z - \Delta_2^z
    \end{pmatrix},
\end{align}
\end{widetext}
where the order of the basis states is $\{\ket{S(2,0)}_0, \ket{\uparrow \uparrow}_0, \ket{\uparrow\downarrow}_0,\ket{\downarrow\uparrow}_0,\ket{\downarrow\downarrow}_0\}$ and we defined the Zeeman vectors as $\boldsymbol \Delta_i = \frac 1 2 \mu_\text{B} \mathbf B \cdot \hat{g}_{\mathrm{Q}i}$.

To simulate the experiment in \figref{fig:fig2}(c) we assume that the ramp in was perfectly adiabatic, i.e., the starting state is a pure state $\ket{\downarrow\downarrow}$, which is the lowest-energy eigenstate of $H_{5\times 5}(\varepsilon_\mathrm{M})$. Note that this assumption neglects the thermalization effect  during the ramp in with the $\ket{\uparrow\downarrow}$ state, which would reduce the contrast in the experiment. After the ramp out, the measurement probability assigned to the $\ket{\downarrow\downarrow}$ state is calculated as
\begin{eqnarray}
    P(\downarrow\downarrow) = \left|\bra{S(2,0)}_0 U(t_\mathrm{ramp})\ket{\downarrow\downarrow}\right|^2 ,
    \label{eq:Pdd_sim}
\end{eqnarray}
where we used $\ket{S(2,0)}_0$ as a measurement projector, and note that using another close-by state that is more appropriate for the description of the latching would only reduce the range of $P(\downarrow\downarrow)$, i.e., the contrast. The propagator in \eqnref{eq:Pdd_sim} is
\begin{eqnarray}
    U(t) = \mathcal T e^{-i\int_0^t dt' H_{5\times 5} (\varepsilon(t'))/\hbar} ,
\end{eqnarray}
with $\varepsilon(t) = \varepsilon_M + (\varepsilon_\mathrm{PSB}-\varepsilon_M)t/t_\mathrm{ramp}$ and where $\mathcal T$ denotes the time ordered integral.

The tunnel couplings $t_\mathrm{o}/h = 0.79 \text{\,GHz}$ and $t_\mathrm{so}/h = 0.125 \text{\,GHz}$ and spin-orbit vector $\textbf{n}_{\mathrm{so}} = (0.81, 0.59, 0)$ were found by generating map plots like the experimental data of \figref{fig:fig2}(c) for different parameter sets, extracting the lines of the characteristic ramp rate given by $P(\downarrow\downarrow) = 1/e$, and performing a least squares minimization to the corresponding lines of the measured data. In this fitting process we disregard the $\phi_\mathbf{B} = 90^\circ$ data from \figref{fig:fig2}(c) due to the absence of corresponding g-factor data. This way, we found significantly better agreement with the measured data and an in-plane spin-orbit vector. Including the data from the $\phi_\mathbf{B} = 90^\circ$ plane strongly violates this expected symmetry.

Finally, the ramp rate used in the Landau-Zener formula is defined as:  
\begin{eqnarray}
    \dv{E_{\mathrm{ST_{-}}}}{\varepsilon} = \frac{E_{\mathrm{ST_{-}}} (\varepsilon_\mathrm{ax})-E_{\mathrm{ST_{-}}} (\varepsilon_\mathrm{ax}-3\Delta_\mathrm{ST})}{3\Delta_\mathrm{ST}}
\end{eqnarray}
where $\varepsilon_\mathrm{ax}$ is the detuning where the $ST_{-}$ gap is minimal. We determined the range of validity for this formula as $E_{\mathrm{ST_{-}}} (\varepsilon_\mathrm{M}) > E_{\mathrm{ST_{-}}} (\varepsilon_\mathrm{ax}-4\Delta_\mathrm{ST})$, which ensures that there is a wide enough dip in the $ST_{-}$ gap such that the transition can be treated as an anticrossing of two energy levels.

\phantom{ABCD} 

\supsection{Theoretical model of the energy spectrum}
\label{app:theoretical model}

In order to account for the finite size of the spin blockade window (i.e., the high energy \zerotwo{} triplet states, as shown in \figref{fig:fig1}(b)), we extend the single-particle Hamiltonian in \eqnref{eq:H1P} with an extra orbital state $\ket{1'}$ on QD1. The modified single-particle Hamiltonian reads
\begin{align}
\begin{split}
        H_{1P'} &= H_{1P} +\left(-\frac \varepsilon 2 + \Delta_\mathrm{orb}\right) \ket{1'}\! \bra{1'}\\
        &\quad+ \frac {\mu_\text{B}} 2 \mathbf B \cdot \hat g_\mathrm{Q1'} \boldsymbol \sigma \ket{1'}\! \bra{1'}\\
        &\quad+ [t_\mathrm{o}' \ket {1'}\! \bra 2 - i t_\mathrm{so}' \mathbf n_\mathrm{so}' \cdot  \boldsymbol \sigma \ket {1'}\! \bra 2 +h.c.],
\end{split}
\label{eq:H1P'}
\end{align}
where we define the orbital splitting in QD1 as $\Delta_\mathrm{orb}$, as well as orbital-specific g-tensor, tunnel couplings and spin-orbit vector. 
This specificity is necessary in full generality to account for the orbital dependence of the tunnel couplings, g-tensors, and so on. 
In the energy spectrum shown in the main text however, we do not distinguish between the $\ket{1}$ and $\ket{1'}$ orbitals in terms of these parameters, as many of these differences (apart from $\Delta_\mathrm{orb}$ which is simply the size of the PSB window) are not known quantitatively.

Using the single-particle Hamiltonian from \eqnref{eq:H1P'} we construct a particle-exchange-symmetric two-particle Hamiltonian
\begin{align}
    H_{2P'} &= H_{1P'} \otimes \mathbb{1} + \mathbb{1} \otimes H_{1P'}
    \label{eq:H2P'} \\
    &\quad + U_H \! \sum \limits_{i,j \in \{1,1'\}}\!\left(\ket i\! \bra i\! \otimes\! \ket j\! \bra j\!  +\! \ket 2\! \bra 2\! \otimes\! \ket 2\! \bra 2\right),
    \notag
\end{align}
where $U_H$ is the Hubbard charging energy. To project this Hamiltonian to the physical subspace, in addition to the five states in \eqnref{eq:basis5a}-\eqref{eq:basis5b} we consider three triplet-like basis states:
\begin{subequations}
    \begin{gather}
        \ket{T_{ss}\zerotwo}_0 = \frac 1 {\sqrt 2} (\ket{1 s} \ket{1' s} - \ket{1' s} \ket{1 s}), \\
        \begin{split}    
            \ket{T_0\zerotwo}_0 &= \frac 1 2 (\ket{1\!\uparrow} \ket{1'\!\downarrow} + \ket{1\!\downarrow} \ket{1'\!\uparrow} \\
            &\qquad -\ket{1'\!\uparrow} \ket{1\!\downarrow} - \ket{1'\!\downarrow} \ket{1\!\uparrow}).
        \end{split}
    \end{gather}
\end{subequations}
We do not consider the singlet-like state of the $1$-$1'$ orbitals because it is even higher in energy due to the single-dot exchange energy.

\end{document}